# Universal temperature scaling of flux line pinning in high-temperature superconducting thin films

**J Albrecht, M Djupmyr, S Brück**

Max-Planck-Institut für Metallforschung, Heisenbergstr. 3, D-70569 Stuttgart, Germany

E-mail: ja@mf.mpg.de

**Abstract**
Dissipation-free current transport in high-temperature superconductors is one of the most crucial properties of this class of materials which is directly related to the effective inhibition of flux line movement by defect structures. In this respect epitaxially grown thin films of $YBa_2Cu_3O_{7-\delta}$ (YBCO) are proving to be the strongest candidates for many widescale applications that are close to realization. We show that the relation between different defect structures and flux line pinning in these films exhibits universal features which are clearly displayed in a detailed analysis of the temperature-dependent behaviour of local critical currents. This allows us to identify different pinning mechanisms at different temperatures to be responsible for the found critical currents. Additionally, the presence of grain boundaries with very low misorientation angles affects the temperature stability of the critical currents which has important consequences for future applications.

Twenty years after the discovery of high-temperature superconductivity in copper oxides by Bednorz and Muller [1] widescale applications of these materials are finally on the horizon [2-5]. Thin epitaxial films, e.g. prepared by pulsed laser deposition (PLD) [6,7], seem to be particularly promising for realizing current carrying tapes, electronic filters and magnetic field sensors [8-10]. A prerequisite for effective operation of all these devices is effective pinning of flux lines in the material to allow the use of high critical current densities and to reduce flux noise. Since not only liquid helium with a boiling point of T=4 K, but also liquid hydrogen (20 K), liquid neon (27 K) and, of course, liquid nitrogen (77 K) are considered as possible coolants, the criterion of effective vortex pinning holds over the whole temperature range of superconductivity. In order to improve the flux pinning capability of thin films of high-temperature superconductors it is crucial to understand the mechanisms that lead to depinning of a flux line. A well established method to obtain insights into flux pinning properties resolves around analysis of the temperature-dependent critical current densities with respect to the local flux line interactions [11-14]. This can be effectively performed by quantitative magnetooptical studies, which allow the direct determination of the magnetic flux density distribution in superconductors and additionally, the local critical current density in the case of a thin film geometry [15]. An example is given in Fig. 1.

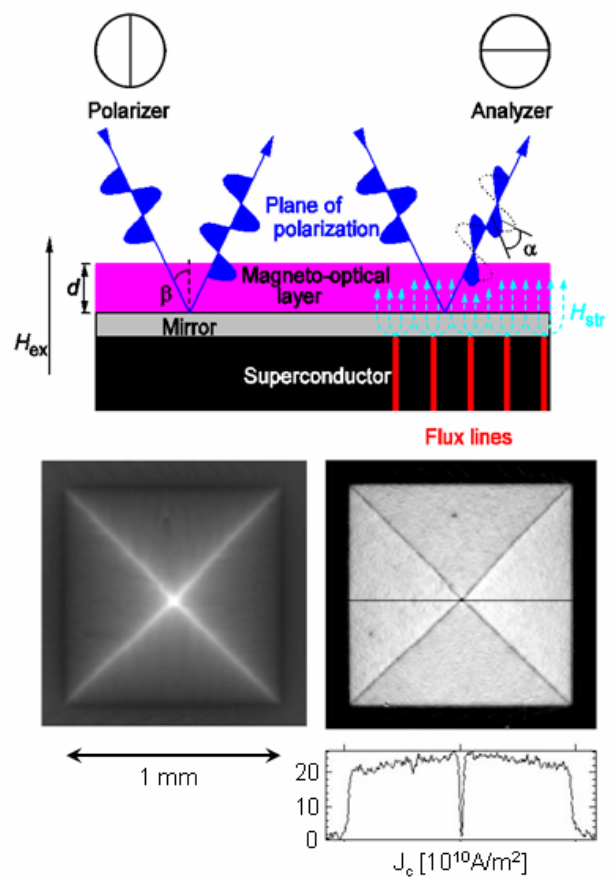

**FIGURE 1.** Sketch of the magnetooptical method (top), the magnetic flux density distribution in the remanent state (bottom left) and the modulus of the critical current density (bottom right) in a square-shaped YBCO film at T=15 K.

With this method, especially by using ferromagnetic iron garnet films as field sensing layers, it is possible to map the local critical current density distribution in thin superconducting films with a spatial resolution of about 5 micrometers. It is determined via a numerical two-dimensional inversion of Biot-Savart´s law which relates the stray field of the sample with the currents flowing:

$$B_z(x,y) = \mu_0 H_{ex} + \mu_0 \int_V \frac{j_x(\mathbf{r}')(y-y') - j_y(\mathbf{r}')(x-x')}{4\pi|\mathbf{r}-\mathbf{r}'|} d^3r'$$

The measured property is the distribution $B_z(x,y)$ which is obtained by magnetooptical images with a subsequent calibration routine. The inversion is then performed in Fourier space using a finite Hanning window for the suppression of high-frequency noise [15]. The application of a self-consistent iteration procedure [16] results in the distribution of local current densities with high accuracy. The obtained current distribution allows the calculation of the integral magnetization that can directly be compared to magnetization measurements which have been performed using a SQUID magnetometer. The magnetization has been measured in the remanent state after having applied an external field of about 0.5 T. This remanent state is subsequently relaxed in increasing temperature. The measurement procedure is exactly equivalent to the MO measurements. A direct comparison of the results is depicted in Fig. 2.

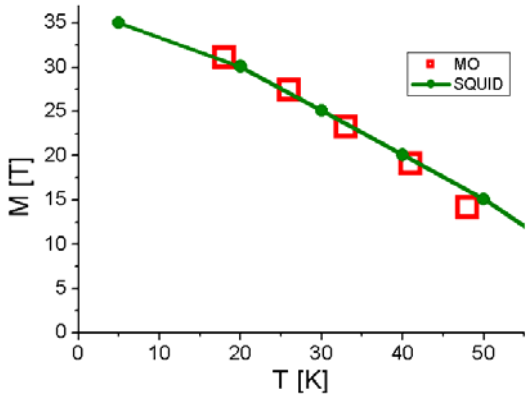

**Figure 2.** The accuracy of the magnetooptical results is demonstrated by calculating the magnetization from the current density distribution and performing a comparison to SQUID results.

Figure 2 demonstrates that the analysis of the magnetooptical images results in a highly accurate determination of current densities.

For a detailed analysis of the current-microstructure relation in thin superconducting films it is not enough to consider integral properties like magnetization values or integrated current densities. The local current density and, thus, the local pinning force on flux lines strongly varies with the local flux density, the local current path and the investigated film geometry [15]. To identify the temperature dependent pinning properties one has to role out these influences by a local analysis at distinguished areas of the superconductor.

This local analysis of the critical current density is performed in the temperature range from T = 7 K to 90 K for optimally doped YBCO films with different microstructure grown by PLD. Details on production and characterization have been published elsewhere [13,14,17]. The results of the local current density analysis are plotted in Fig. 3 from bottom to top for an individual low-angle grain boundary (misorientation angle 3°, reverse triangles), a standard high-quality YBCO film (which was also used for the Figures 1 and 2, triangles), a YBCO film with improved growth island connectivity (circles) and a step flow grown YBCO film with nearly no grain boundaries along the current direction (squares). At this point we have to state that the high current densities found in all of these samples can only be explained in a strong pinning regime and by the absence of weak links.

Additionally, it is necessary to state that in case of the individual 3° tilt grain boundary strong flux focusing leads to a substantial flux line interaction in and in vicinity of the boundary. This leads to an enhancement of the critical current density due to magnetic dipolar interaction of the grain boundary vortices with the adjacent Abrikosov vortices in the banks. To extract the intrinsic pinning of the grain boundary flux lines, this interaction, which is additive, has to be subtracted. For small flux densities this additional pinning can be modelled by two-body interactions of flux lines [17] and subtracted from the current density data. For all other samples considered, like the one depicted in Fig. 1, the $j(B)$ dependence is rather weak, because all vortices exhibit an equivalent pinning behaviour. So, the current density data can be extracted in the remanent state in areas of the superconductor where the local flux density changes its sign. Here, the contribution of vortex interactions can be neglected.

Note, that it is absolutely necessary to take care on the role of the local flux density which is easily demonstrated by comparing the results depicted in Figure 2 and 3, respectively.

Considering the obtained *local* current densities it is found that at temperatures above $T$=40 K all films show the same linear temperature scaling of $j_c$ with respect to the reduced temperature $t=(1-T/T_c)$. Below a crossover region around 30-40 K the scaling changes and different slopes are observed in the log($j_c$)–log(t) plot.

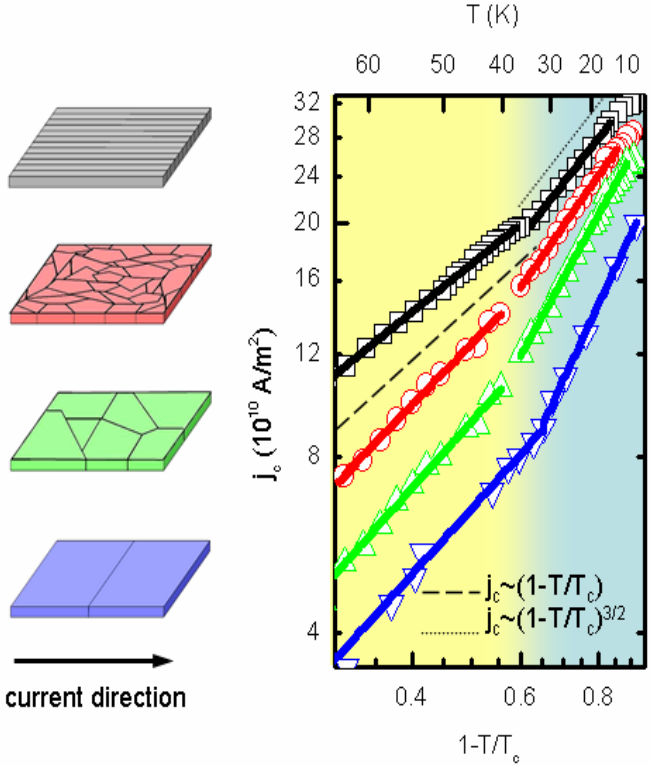

**Figure 3.** From bottom to top the data are plotted for an individual 3° tilt grain boundary (reverse triangles), a regularly grown epitaxial film on a flat substrate (triangles), a film on a pre-patterned substrate with improved coalescence regions (circles) and a step-flow grown film on a vicinal cut substrate with nearly no grain boundaries along the current direction (squares). The dashed line represents a linear relation $j \sim (1-T/T_c)$ and the dotted line a power-law $j \sim (1-T/T_c)^{3/2}$. $T_c$ is around T=90 K for all films.

The observed temperature scaling allows one to identify different vortex depinning mechanisms at different temperatures. Above $T=40$ K, we find, in all cases, a linear behaviour $j_c \sim (1-T/T_c)$, exactly the linear temperature dependence of the flux line self energy $\varepsilon(T)$ given by Ginzburg-Landau theory, which holds quite well in this temperature region. The pinning energy $U_0$, which denotes the depth of a potential well at zero current, is proportional to the flux line energy $\varepsilon(T)$. This is only correct for individual, non-interacting vortices, but this condition is quite well fulfilled in our measurement technique provided, we confine ourselves to current density data in areas of low magnetic flux densities. The observation of a linear temperature dependence of $j_c$ for **all** samples above $T=40$ K leads one to the conclusion that the critical current density is proportional to the activation energy $U_0$. This allows us to consider thermally activated depinning as the limiting factor for the critical current in YBCO films in this temperature range, because the thermally activated process at finite current densities has to overcome a small barrier $U(j)$ which is proportional to $U_0(T)$ [18] leading to $j_c \sim (1-T/T_c)$. The offset of the $j_c$ values for the investigated samples finds its origin in different depths of the pinning potential well.

In the temperature region below a crossover range between 30 and 40 K, the rate of thermally activated depinning goes down [19]. The thermal energy, $kT$, is not sufficient for substantial thermally activated relaxation. Here, the shape of the individual pinning potential comes into play. The force that is required to remove a flux line from a pinning site is given by the gradient of the energy landscape in the direction of the acting Lorentz force. In high-temperature superconductors the dominant contribution to this energy gradient comes from flux line core pinning, i.e. the variation of the energy of the normal conducting vortex core on the background of variations of the superconducting order parameter in the material. In the case of core pinning one can assume the following expression for the elastic energy of a flux line displaced by a distance $u$ from an optimal pinning position [20,21]

$$\varepsilon_{pin}(u) = -\frac{\varepsilon_0}{2} \frac{r_0^2}{u^2 + 2\xi^2} \quad (1)$$

where $r_0$ is the radius of a pinning site and $\xi$ is the superconducting coherence length. The derivative of equation (1) with respect to $u$ directly yields the pinning force

$$f_{pin}(u) = \varepsilon_0 \frac{u r_0^2}{(u^2 + 2\xi^2)^2} \quad (2)$$

At very low local flux densities flux lines usually find the most effective pinning sites, which are normal conducting or insulating regions of size of the flux line core or smaller. Owing to a finite proximity effect the area of suppressed order parameter cannot be smaller than the flux line core, which allows us to replace $r_0$ by $\xi$. Taking now the maximum of equation (2) at $u=(2/3)^{1/2}\xi$ and replacing $r_0$ one ends up with:

$$j_c(T) = \frac{1}{\Phi_0} f_{pin}^{max} \propto \Phi_0 \frac{\varepsilon_0}{\xi} \propto (1-\frac{T}{T_c})^{3/2}$$

where $\Phi_0$ denotes the flux quantum $\Phi_0=2.07\times10^{-15}$ Tm$^2$.

The exponent of $s=3/2$, which results is exactly that found (within error bars) for the black data points of Fig. 2 at temperatures between 15 and 40 K.

Owing to the different microstructure the power-law with an exponent of $s=3/2$ can not be detected in all the other samples. These samples exhibit two-dimensional defect structures perturbing the path of the supercurrents, which are the coalescence regions between the growth islands in case of the samples grown on flat and irradiated substrates and the individual grain boundary. Analyzing the data depicted in Fig. 2, we find exponents of *s=1.7* for the film with highly connected growth islands, *s=2.0* for the standard YBCO film and *s=2.5* for the individual 3° grain boundary. This systematic increase of the exponent s clearly shows that the temperature dependence of the critical current density is strongly affected by the quality of low-angle grain boundaries in the films, even if the misorientiation angle is of the order of 1° (for the standard film) or below (for the film on an irradiated substrate). This is surprising because in transport measurements at low temperatures one finds a significant suppression of the critical current density only for misorientation angles of 4-5° and above [22,23]. Note that local magnetization experiments show a slightly smaller critical angle for the current limiting role of grain boundaries of 2-3° [17], but in transport experiments the vortices are stabilized by the self field of the supercurrents, which is the reason for the difference.

As a result of this analysis we can state that networks of low-angle grain boundaries as found in thin films containing growth islands clearly affect the temperature stability of high critical currents found at low temperatures. This is strongly supported by the data of the individual 3° grain boundary, which exhibits the discussed characteristics in a more pronounced way. In all samples the critical current density $j_c$ is not directly limited by the depairing current $j_0$ across the grain boundaries. For angles lower than 5° $j_0$ is always larger than $j_c$. Only above 5° the exponential decrease of $j_0$ sets an upper limit for $j_c$ [22]. For the considered samples, a slight decrease of $j_0$ leads to an anisotropic deformation of the vortices along the grain boundary [24] and, thus, to a decrease of the pinning forces.

For the low temperature region below *T*=15 K, a power-law description of the data in Fig. 2 fails. This is due to the onset of quantum creep of flux lines, which was found to be substantial in YBCO below *T*=15 K [25].

Considering now all current-temperature relations displayed in Fig. 2 there is a clear correlation between the magnitude and the low-temperature slope of the $j_c$-curves and the film microstructure. The samples can be classified by the different kind of interfaces the supercurrents have to cross.

It is now possible to present a phase diagram-like cartoon for the current limiting factors in strong pinning thin films of high-temperature superconductors in small magnetic fields which is depicted in Fig. 3.

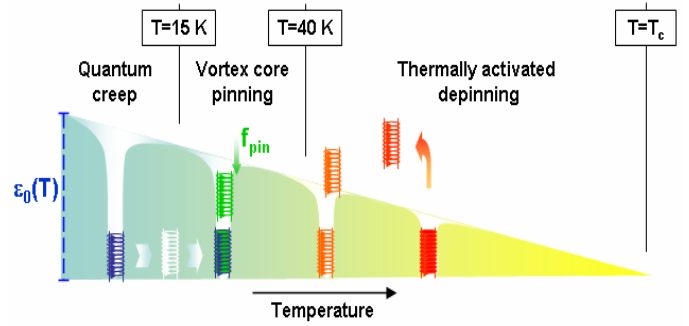

**Figure 4.** The experimental results can be summarized in a phase diagram-like cartoon for low magnetic flux densities. Three different generic pinning scenarios are found for all investigated YBCO thin films. However, the transitions are not sharp and are best described by cross-over regions.

The sketch shows the temperature increasing to the right. This is also indicated by colours from blue to yellow referring to temperatures from *T*=7 K to the superconducting transition temperature which is around 90 K for all films. An analysis of the temperature-dependent data presented in Fig. 2 allows us to directly distinguish between the three regimes. This holds for all the different superconducting films investigated in this work which have in common that there are no weak links along the current path.

The first regime up to a temperature of *T*=15 K exhibiting no power-law behaviour of $j_c(T)$ is governed by quantum creep of flux lines between different pinning sites. Above T=15 K up to a temperature between 35 and 40 K we conclude that effective flux line core pinning is responsible for the high current densities found. In this temperature regime the current suppressing influence of low-angle grain boundaries becomes particular substantial. The finite transparency of the grain boundaries locally suppresses the depairing current density $j_0(T)$ which modifies the current distribution and, thus, the energy of a flux line. This leads to a larger exponent, which means a stronger decrease of $j_c$ with temperature. Above a transition region between 35 and 40 K all samples show a linear decrease of $j_c$ with temperature, indicating the energetic depth of the pinning sites as decisive factor where a flux line is depinned by thermal activation. It is important to note that the pinning sites of the flux lines need not to be the same in the different regimes. It is more likely that at low temperatures small and narrow defects are the most effective pinning sites which are just smeared out if enough thermal energy is provided.

In summary, we have identified the mechanisms of flux line pinning that are responsible for the high critical current densities of epitaxial YBCO films. This result can only be obtained by the consideration of local current density data. The overall influence of geometry and flux density effects present in globally obtained data can not reveal the temperature-dependent pinning properties [26]. From our local results we can state that low-angle grain boundaries with misorientation angles of the order of 1° do not affect the critical currents at the temperature of liquid helium but definitely suppress the current values at higher temperatures which are interesting for realizing cooling systems consisting of liquid hydrogen or neon. This substantial drop of $j_c$ with temperature results in large differences of the critical currents at higher temperatures above $T$=40 K, where thermally activated depinning of flux lines limits the critical current and leads to a universal temperature scaling of the critical currents in all considered samples with strong pinning.

## Acknowledgments

The authors are grateful to S. Bending, E. H. Brandt and H.-U. Habermeier for helpful discussions and to G. Cristiani and H.-U. Habermeier for the preparation of the excellent YBCO films.